\documentclass[conference]{IEEEtran}
\IEEEoverridecommandlockouts

\usepackage{cite}
\usepackage{amsmath,amssymb,amsfonts}
\usepackage{algorithmic}
\usepackage{graphicx}
\usepackage{textcomp}
\usepackage{xcolor}
\usepackage[ruled,linesnumbered]{algorithm2e}
\DeclareMathOperator{\UEdk}{\mathrm{UE}_{\mathtt{k}}}
\DeclareMathOperator{\SSEk}{\mathtt{SSE}_k}

\def\BibTeX{{\rm B\kern-.05em{\sc i\kern-.025em b}\kern-.08em
    T\kern-.1667em\lower.7ex\hbox{E}\kern-.125emX}}
    
\usepackage[margin=15mm,top=18.1mm,bottom=26mm]{geometry}

\begin{document}

\title{Tri-Hybrid Beamforming Design for DMA-Aided Secure ISAC Systems \\
\thanks{This work was supported by the U.K. Engineering and Physical Sciences Research Council (EPSRC) grant (EP/X04047X/2) for TITAN Telecoms Hub. The work of J. He was supported by the HORIZON-MSCA through the project ECO-ISAC under Grant 101274088. The work of H.~Q.~Ngo was supported by a research grant from the Department for the Economy Northern Ireland under the US-Ireland R\&D Partnership Programme. The work of M. Matthaiou was supported by the European Research Council (ERC) under the European Union’s Horizon 2020 Research and Innovation Programme (grant agreement No. 101001331).}
}

\author{Siyi Li\textsuperscript{*\dag}, Zhuoming Li\textsuperscript{*}, Mohammadali Mohammadi\textsuperscript{\dag}, Jiajun He\textsuperscript{\dag}, Hien Quoc Ngo\textsuperscript{\dag}, and Michail Matthaiou\textsuperscript{\dag}\\ \textsuperscript{*}School of Electronics and Information Engineering, Harbin Institute of Technology, Harbin, China.\\
\textsuperscript{\dag}Centre for Wireless Innovation (CWI), Queen’s University Belfast, U.K.
\\Email: lisiyi1103@stu.hit.edu.cn, zhuoming@hit.edu.cn, \{m.mohammadi, j.he, hien.ngo, m.matthaiou\}@qub.ac.uk
}

\maketitle

\begin{abstract}
This paper proposes a tri-hybrid beamforming scheme for secure integrated sensing and communication (ISAC) with a dynamic metasurface antenna (DMA) architecture, where the base station (BS) is capable of communicating with legitimate users and sensing the target. There is also an eavesdropper in the system intending to eavesdrop on the confidential information. The tri-hybrid beamforming design problem is formulated with the objective of maximizing the sensing signal-to-noise ratio (SNR) under the constraints of secrecy spectral efficiency (SSE), transmit power, and physical structure limitations. We first solve the problem and obtain the optimized fully-digital beamforming solution through successive convex approximation (SCA) and semidefinite relaxation (SDR) approaches. A triple alternating optimization scheme is then developed to iteratively optimize the digital, analog, and DMA beamformers, progressively approximating the fully-digital solution. Numerical results demonstrate that the proposed secure tri-hybrid beamforming design for DMA-aided ISAC improves the sensing SNR by approximately 3 dB compared to a tri-hybrid beamforming scheme with a fixed DMA electromagnetic design.
\end{abstract}

\vspace{0.5em}
\begin{IEEEkeywords}
Dynamic metasurface antenna, integrated sensing and communication, physical layer security, tri-hybrid beamforming. 
\end{IEEEkeywords}

\section{Introduction}
\IEEEPARstart{I}{SAC} is capable of unifying sensing and communication functionalities within a single system, and hence is regarded as a promising paradigm for next-generation mobile networks \cite{b1}. However, achieving robust sensing and communication performance typically requires large-scale antenna arrays \cite{b2}, where conventional fully-digital architectures introduce extremely high power consumption and suffer from limitations in size and deployment flexibility. To address these challenges, the \textit{tri-hybrid beamforming architecture} has recently been introduced as a promising solution, enabling enhanced design flexibility by exploiting an additional antenna layer \cite{b3,tri}. 
\par The tri-hybrid beamforming architecture can be realized using reconfigurable antenna arrays, including pixel antennas \cite{b4}, radiation-center reconfigurable antennas (RCRAs) \cite{b5}, and DMAs \cite{b6}. The authors in \cite{b4} designed a pixel antenna-assisted tri-hybrid beamforming and maximized the system sum rate under practical constraints, while the authors in \cite{b5} designed the tri-hybrid beamforming assisted by an RCRA to maximize energy efficiency (EE). The tri-hybrid beamforming design for DMA-assisted ISAC systems was studied in \cite{b6}, where the authors proposed a joint optimization framework aiming to achieve a trade-off between communication and sensing SNR performance.
\par On the other hand, security remains a critical and challenging issue in ISAC systems. Since sensing requires directing beams toward targets, it inherently increases the risk of information leakage to potential eavesdroppers (Eves) \cite{b7}. Physical layer security (PLS) technologies can exploit the characteristics of wireless channels to enhance security performance. Furthermore, in ISAC systems, dedicated sensing waveforms, that do not carry communication information, can be further designed to generate interference toward an Eve, thereby improving the security performance \cite{b8}. Existing works have investigated PLS design in ISAC systems under a hybrid beamforming architecture \cite{b9,b10}. Specifically, the authors in \cite{b9} investigated how many sensing beams, serving both target sensing and Eve jamming, are required in a hybrid architecture to achieve secure communication. In \cite{b10}, the security of a sub-connected active reconfigurable intelligent surface (SC-ARIS) ISAC system was studied, where hybrid and SC-ARIS beamforming schemes were jointly optimized using alternating optimization. 
\par Most existing studies either focus on tri-hybrid beamforming design to improve communication performance, or concentrate on PLS design under conventional hybrid analog-digital beamforming architectures. To our knowledge, the design of tri-hybrid beamforming for secure ISAC systems has not yet been investigated. Motivated by this observation, we propose a tri-hybrid beamforming design for DMA-aided secure ISAC systems. The key contributions of our work can be summarized as follows:
\begin{itemize}
\item  We establish a secure tri-hybrid ISAC system model assisted by DMA and formulate an optimization problem that maximizes sensing SNR under SSE, transmit power and physical structure constraints. The Cramér-Rao bound (CRB) is also derived to evaluate the system performance.
\item We propose a triple alternating optimization algorithm to jointly design the tri-hybrid beamformer to approximate the fully-digital beamformer. The fully-digital beamformer is obtained using SDR and SCA, followed by the alternating optimization of the digital, analog and DMA beamformers for the tri-hybrid architecture.
\item Numerical results demonstrate that, compared with conventional hybrid beamforming under the same antenna aperture and the tri-hybrid beamforming with fixed DMA design, the proposed secure tri-hybrid beamforming design achieves higher sensing SNR and improved CRB performance under the same SSE requirement.
\end{itemize}
\emph{Notations}: Bold lowercase (uppercase) letters denote vectors (matrices); the conjugate transpose, transpose, inverse, rank and trace operators are denoted by $(\cdot)^\text{H}$, $(\cdot)^\text{T}$, $(\cdot)^{-1}$, ${\rm rank}(\cdot)$ and ${\rm Tr}(\cdot)$, respectively; $\left\Vert \cdot \right\Vert_F$ denotes the Frobenius norm; $\otimes$ stands for Kronecker product; $\Re\{\cdot\}$ is the real part of a complex entry; $\rm{blkdiag}(\cdot)$ is the block-diagonal matrix operator; $\left[ \cdot \right]^+$ denotes ${\rm max} \left\{ \cdot,0 \right\}$; Finally, $\mathcal{CN}(\boldsymbol{\mu},\boldsymbol{\Sigma})$ denotes the complex Gaussian vector with mean $\boldsymbol{\mu}$ and covariance $\boldsymbol{\Sigma}$. 

\section{System Model and Problem Formulation}
\subsection{Tri-Hybrid Beamforming Model}
\par Consider a secure ISAC system, as shown in Fig. 1, where the BS simultaneously transmits communication signals to $K$ legitimate user equipment (UEs) and senses a designated target. Meanwhile, a potential Eve attempts to intercept the confidential information intended for the UEs. The BS adopts a DMA-assisted tri-hybrid beamforming architecture. Specifically, the DMA consists of $N_w$ waveguides, and each waveguide has $N_r$ reconfigurable radiating elements. The total number of radiating elements at the BS is: $N=N_{w}N_{r}$. 

\par The tri-hybrid architecture at the BS includes digital, analog, and additional DMA beamformers. Let $\mathbf{F}_{\mathrm{dig}}\in \mathbb{C}^{N_{RF}\times (K+1)}$ be the digital beamformer, which maps the baseband signals to the $N_{RF}$ radio frequency (RF) chains. The analog beamformer is denoted by $\mathbf{F}_{\mathrm{RF}} \in \mathbb{C}^{N_w\times N_{RF}}$, which maps the RF chain outputs to the antenna layer \cite{b11}. The analog beamformer is implemented using a fully-connected architecture, 
where each RF chain is connected to all waveguides through phase shifters. Due to the hardware limitations of phase-only control, the elements in the analog beamformer are subject to the constant modulus constraint, which can be expressed as $\left|[\mathbf{F}_{\mathrm{RF}}]_{m,n}\right|=1/\sqrt{N_w}$.
\par Let $\mathbf{F}_{\mathrm{dma}} \in \mathbb{C}^{N\times N_w}$ be the DMA beamformer. Due to the fact that all $N_r$ radiating elements within each waveguide are powered by a common feeding port, the entire antenna array can be regarded as $N_w$ groups of subarrays. As a result, the DMA beamformer admits a block diagonal-form, as follows
\begin{equation}\label{Fdma}
\mathbf{F}_{\mathrm{dma}} = \mathrm{blkdiag}(\mathbf{m}_1, \mathbf{m}_2, \dots, \mathbf{m}_{N_w}),
\end{equation}
where $\mathbf{m}_i=[m_{i,1},m_{i,2},\cdots m_{i,N_r}]^T \in \mathbb{C}^{N_r \times 1}$ denotes the DMA coefficient vector associated with the $i$-th waveguide. Specifically, the $\ell$-th element of $\mathbf{m}_i$ is modeled as
\begin{equation}\label{mil}
m_{i,\ell} = q_{i,\ell} \, \alpha_{i,\ell},
\end{equation}
where $q_{i,\ell} = e^{-(\alpha + j\beta)d_{i,\ell}}$ is the waveguide propagation coefficient, with $\alpha$ and $\beta$ denoting the attenuation coefficient and wavenumber, respectively; $d_{i,\ell}$ is the propagation distance from the input to the $\ell$-th radiating element. Due to the physical structure constraint of the DMA, the coefficient $\alpha_{i,\ell}$, which characterizes the tunable response of each radiating element, exhibits an inherent coupling between the amplitude and phase. This response is governed by a Lorentzian-type constraint as \cite{b6}
\begin{equation}\label{alphail}
\alpha_{i,\ell} = \frac{1}{2}\left(j + e^{j\theta_{i,\ell}}\right),
\end{equation}
where $\theta_{i,\ell} \in [0, 2\pi]$ is the configurable phase shift.

  \begin{figure}[!t]
    \centering
    \includegraphics[width=3.4in]{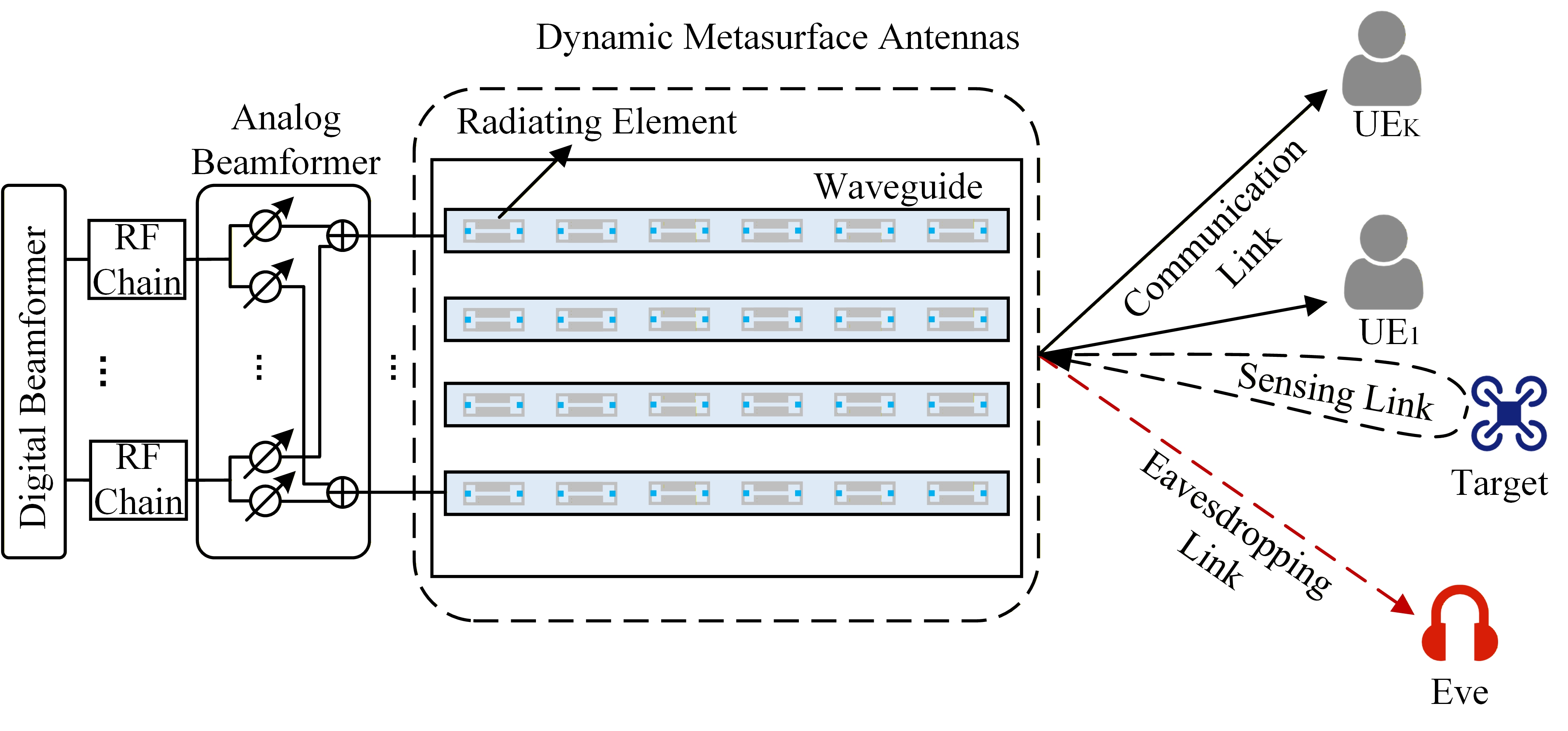}
    \caption{Secure tri-hybrid ISAC system model.}
    \vspace{-1em}
    \label{fig1}
\end{figure}

\par Based on the tri-hybrid beamforming architecture, the signal transmitted by the BS is given by
\begin{equation}\label{yk}    \mathbf{x}=\mathbf{F}_{\mathrm{dma}}\mathbf{F}_{\mathrm{RF}}\sum\nolimits_{k=0}^K\mathbf{f}_{\mathrm{dig}_k}s_k(t),
\end{equation}
where $\mathbf{f}_{\mathrm{dig}_0}\in \mathbb{C}^{N_{RF}\times 1}$ and $\mathbf{f}_{\mathrm{dig}_k}\in \mathbb{C}^{N_{RF}\times 1}$ are the first and the $(k+1)$-th columns of $\mathbf{F}_{\mathrm{dig}}$ representing the digital beamformers for sensing and for communicating with legitimate UE $k$ (denoted by $\UEdk$), receptively; $s_0(t)$ is the unit power sensing symbol and $s_k(t)\sim\mathcal{CN}(0,1)$, $k=1,\ldots,K$, denotes the communication symbol for $\UEdk$.
Thus, the signal received at   $\UEdk$ can be expressed as
\begin{equation}\label{yk}    y_k(t)=\mathbf{g}_k^H\mathbf{F}_{\mathrm{dma}}\mathbf{F}_{\mathrm{RF}}\sum\nolimits_{k=0}^K\mathbf{f}_{\mathrm{dig}_k}s_k(t)+n_{k}(t),
\end{equation}
where $\mathbf{g}_k \in \mathbb{C}^{N\times1}$ is the millimeter-wave (mmWave) channel from the BS to $\UEdk$ modeled by the sparse geometric multi-path model, and $n_k(t)\sim\mathcal{CN}(0,\sigma_k^2)$ is the zero-mean additive white Gaussian noise (AWGN) with variance $\sigma_k^2$ at $\UEdk$.
\par The signal-to-interference-plus-noise ratio (SINR) at $\UEdk$ is then:
\begin{equation}\label{gamma_k}
    \gamma_k = \frac{|\mathbf{g}_k^H\mathbf{F}_{\mathrm{dma}}\mathbf{F}_{\mathrm{RF}}\mathbf{f}_{\mathrm{dig}_k}|^2}{\sum\limits_{j\neq k}^K|\mathbf{g}_k^H\mathbf{F}_{\mathrm{dma}}\mathbf{F}_{\mathrm{RF}}\mathbf{f}_{\mathrm{dig}_j}|^2+\left|\mathbf{g}_k^H\mathbf{F}_{\mathrm{dma}}\mathbf{F}_{\mathrm{RF}}\mathbf{f}_{\mathrm{dig}_0}\right|^2+\sigma_k^2},
\end{equation}
\par Similarly, the SINR at  Eve when intercepting $\UEdk$ is
\begin{equation}\label{gamma_ek}
    \gamma_{k}^e = \frac{|\mathbf{g}_e^H\mathbf{F}_{\mathrm{dma}}\mathbf{F}_{\mathrm{RF}}\mathbf{f}_{\mathrm{dig}_k}|^2}{\sum\limits_{j\neq k}^K|\mathbf{g}_e^H\mathbf{F}_{\mathrm{dma}}\mathbf{F}_{\mathrm{RF}}\mathbf{f}_{\mathrm{dig}_j}|^2+\left|\mathbf{g}_e^H\mathbf{F}_{\mathrm{dma}}\mathbf{F}_{\mathrm{RF}}\mathbf{f}_{\mathrm{dig}_0}\right|^2+\sigma_e^2},
\end{equation}
where $\mathbf{g}_e \in \mathbb{C}^{N\times1}$ is the channel  from the BS to Eve, $\sigma_e^2$ is the variance of the AWGN at Eve. Hence, the SSE at $\UEdk$ is:
\begin{equation}\label{SSE}
 \SSEk=\left[ {\rm log_2}\left(1+ \mathrm{\gamma}_k\right)-{\rm log_2}\left(1+ \mathrm{\gamma}_{k}^e\right)\right]^+.
 \end{equation}
\par For the sensing part, we assume that the DMA is arranged in a uniform planar array (UPA) with $N$ elements \cite{b6}. Let $\theta$ and $\phi$ denote the elevation angle and the azimuth of the target, respectively, the steering vector can be expressed as $\mathbf{a}(\theta,\phi) = \mathbf{a}_h(\theta,\phi) \otimes \mathbf{a}_v(\theta)$ with
\begin{equation}
\mathbf{a}_h(\theta,\phi) = \left[1, e^{-j\frac{2\pi h}{\lambda}\sin\theta\sin\phi}, \dots, e^{-j\frac{2\pi h}{\lambda}(N_r-1)\sin\theta\sin\phi} \right]^T,
\end{equation}
\begin{equation}
\mathbf{a}_v(\theta) = \left[1, e^{-j\frac{2\pi v}{\lambda}\cos\theta}, \dots, e^{-j\frac{2\pi v}{\lambda}(N_w-1)\cos\theta} \right]^T,
\end{equation}
where $h$ and $v$ denote the inter-element spacing in the horizontal and vertical directions, respectively. The construction of $\mathbf{a}_t(\theta,\phi)$ and $\mathbf{a}_r(\theta,\phi)$ is the same as that of $\mathbf{a}(\theta,\phi)$. For notational simplicity, we omit the expression on $\theta$ and $\phi$ in the following, and use $\mathbf{a}$ to denote $\mathbf{a}(\theta,\phi)$. 
\par Assuming a point target with the complex coefficient $\beta_t$ including two-way pathloss coefficient and the Radar Cross Section (RCS), the received echo signal at the BS can be modeled as
\begin{equation}
\mathbf{y}_s(t) = \beta_t\mathbf{a}_r\mathbf{a}_t^H\mathbf{F}_{\mathrm{dma}}\mathbf{F}_{\mathrm{RF}}\sum\nolimits_{k=0}^K\mathbf{f}_{\mathrm{dig}_k}s_k(t)+ \mathbf{n}_s(t),
\end{equation}
where $\mathbf{n}_s \sim \mathcal{CN}(\mathbf{0}, \sigma_s^2 \mathbf{I})$ is the AWGN.
\par With maximum-ratio combining, i.e., $\mathbf{y}_s(t)$ is projected onto $\mathbf{a}_r^H$, the sensing SNR is
\begin{equation}
\gamma_s=\frac{|\beta_t|^2N\mathbf{a}_t^H\mathbf{F}_{\mathrm{dma}}\mathbf{F}_{\mathrm{RF}}\sum_{k=0}^K\mathbf{f}_{\mathrm{dig}_k}\mathbf{f}^H_{\mathrm{dig}_k}\mathbf{F}_{\mathrm{RF}}^H\mathbf{F}_{\mathrm{dma}}^H\mathbf{a}_t}{\sigma_s^2}.
\end{equation}
\par For the sensing part, the CRB is also adopted as a performance metric to evaluate the sensing performance. The Fisher information matrix (FIM) of estimating $\theta$ and $\phi$ can be expressed as \cite{bCRLB}
\begin{equation}\label{fim}  \mathbf{F}=\begin{bmatrix}F_{\theta\theta}&F_{\theta\phi}\\ F_{\theta\phi}^{\mathrm{T}}& F_{\phi\phi}\end{bmatrix},
\end{equation}
where each element of $\mathbf{F}$ can be expressed as $F_{ \boldsymbol{\zeta}_i\boldsymbol{\zeta}_j}=\frac{2}{\sigma_s^2}\Re\Big\{\frac{\partial (\mathbf{y}_s)^{H}}{\partial\boldsymbol{\zeta}_i}\frac{\partial(\mathbf{y}_s)}{\partial\boldsymbol{\zeta}_j}\Big\}$ with $\boldsymbol{\zeta}=\big\{\theta,\phi\big\}$. By substituting the expression of the received echo signal into \eqref{fim} and performing further simplification, each entry of the FIM is given by
\begin{equation}
 F_{ \boldsymbol{\zeta}_i\boldsymbol{\zeta}_j}=\frac{2L|\beta_t|^2}{\sigma_s^2}\Re\big\{{\rm Tr}(\mathbf{A}_{\boldsymbol{\zeta}_i}\mathbf{F}_{\mathrm{eff}}\mathbf{A}_{\boldsymbol{\zeta}_j}^H)\big\}, \label{F1}
\end{equation}
where $L$ is the time-domain snapshot number; $\mathbf{A}_{\boldsymbol{\zeta}_i}=\frac{\partial\mathbf{a}_r}{\partial\boldsymbol{\zeta}_i}\mathbf{a}_t^H+\mathbf{a}_r\frac{\partial\mathbf{a}_t^H}{\partial\boldsymbol{\zeta}_i}$, and $\mathbf{F}_{\mathrm{eff}}=\mathbf{F}_{\mathrm{dma}}\mathbf{F}_{\mathrm{RF}}\big(\sum\nolimits_{k=0}^K\mathbf{f}_{\mathrm{dig}_k}\mathbf{f}^H_{\mathrm{dig}_k}\big)\mathbf{F}_{\mathrm{RF}}^H\mathbf{F}_{\mathrm{dma}}^H$.
\par We adopt the trace of the CRB matrix $\mathbf{C}$ as the scalar performance metric for estimating $\theta$ and $\phi$, which can be expressed as
\begin{equation}
{\rm CRB}={\rm Tr}(\mathbf{C}) = {\rm Tr}({\mathbf{F}}^{-1}).
\end{equation}

\subsection{Problem Formulation}
\par Based on the tri-hybrid beamforming architecture, we jointly design the DMA beamformer, the analog beamformer and the digital beamformer. The tri-hybrid beamforming design aims to maximize the sensing SNR while satisfying several practical constraints, including the constraints on the hardware structures of the analog beamformer and the DMA beamformer, the SSE performance and the total transmit power. The corresponding optimization problem can be formulated as
\begin{subequations}\label{P1}
\begin{alignat}{2}
\mathcal{P}_1:\hspace{-1.3em}
& & &\hspace{1em}\max_{\mathbf{F}_{\mathrm{dma}}, \mathbf{F}_{\mathrm{RF}}, \mathbf{F}_{\mathrm{dig}}}
  \gamma_s  \label{P1:obj}\\
& & &\hspace{3.75em}\text{s.t.} \hspace{2em} \SSEk \geq R_{sk},\label{P1:SSE}\\
& & &\hspace{7em}{\rm Tr}\big(\mathbf{F}_{\mathrm{eff}}\big) \leq P, \label{P1:P}\\
& & &\hspace{7em}\left|[\mathbf{F}_{\mathrm{RF}}]_{m,n}\right|=1/\sqrt{N_w},\label{P1:f}\\
&\;& &\hspace{7em} \eqref{Fdma}-\eqref{alphail},\label{P1:dma}
\end{alignat}
\end{subequations}
where $P$ is the maximum transmit power at the BS, and $R_{sk}$ is the minimum SSE requirement. It is seen that problem $\mathcal{P}_1$ is a complex problem with coupled optimization variables, including 1) a non-convex SSE constraint, 2) the constant-modulus constraint \eqref{P1:f} on the analog beamformer, and 3) the structural constraint \eqref{P1:dma} of the DMA. As a result, problem $\mathcal{P}_1$ cannot be solved directly. To address this issue, an efficient algorithm is developed in the following section.

\section{Tri-Hybrid Beamforming Design}
\par To solve problem $\mathcal{P}_1$, we consider a triple alternating algorithm, where the three beamformers are designed iteratively. We first consider the equivalent overall beamformer formed by the tri-hybrid architecture. By relaxing the hardware constraints imposed by the analog and DMA beamforming structures, this equivalent beamformer can be freely optimized, leading to a fully-digital beamforming design. This design can be regarded as an architecture where each antenna is connected to its own RF chain. The resulting fully-digital beamformer is then used as the target solution, which the tri-hybrid beamforming design aims to approximate. Therefore, we first define $\mathbf{W}_k=\mathbf{F}_{\mathrm{dma}}\mathbf{F}_{\mathrm{RF}}\mathbf{f}_{\mathrm{dig}_k}\mathbf{f}^H_{\mathrm{dig}_k}\mathbf{F}_{\mathrm{RF}}^H\mathbf{F}_{\mathrm{dma}}^H=\mathbf{w}_k\mathbf{w}_k^H$ to represent the overall beamforming process as a unified transformation. Note that $\mathbf{w}_k$ can be regarded as the fully-digital beamforming solution for problem $\mathcal{P}_1$, such that the problem for $\mathbf{W}_k$ can be equivalently expressed as
\begin{subequations}\label{P1.1}
\begin{alignat}{2}
\mathcal{P}_{1.1}:\;
&\max_{\mathbf{W}_k}
&\;&\quad  \mathbf{a}_t^H\Big(\sum\nolimits_{k=0}^K\mathbf{W}_k\Big)\mathbf{a}_t\label{P1.1:obj}\\
&\hspace{1em}\text{s.t.}\;& & \hspace{1em} {\rm rank}(\mathbf{W}_k)=1,\;\;\mathbf{W}_k \succeq \mathbf{0},\\
& & &\hspace{1em}\eqref{P1:SSE},\;\eqref{P1:P}.\label{P1.1:c}
\end{alignat}
\end{subequations}
\par However, it is seen in \eqref{P1.1} that problem $\mathcal{P}_{1.1}$ is still challenging to solve due to the rank-1 and the SSE constraints. Thus, we first utilize the SDR method to relax the rank-1 constraint \cite{SDR}. Then, to tackle the SSE constraint, a set of auxiliary variables $\big\{\{\tau_k\}_{k=1}^K,\{\zeta_k\}_{k=1}^K,\tau_e,\zeta_e\big\}$ is introduced to decompose the SSE constraint into separate constraints for the legitimate UEs and the Eve. By utilizing the SCA approach, these constraints are further transformed, while the auxiliary variables are constrained to satisfy the threshold of the original SSE constraint.
\par By substituting \eqref{gamma_k} and \eqref{gamma_ek} into \eqref{SSE}, the SSE constraint for $\UEdk$ can be rewritten as
\begin{subequations}
\begin{align} &\mathbf{g}_k^H\big(\sum\nolimits_{k=0}^K\mathbf{W}_k\big)\mathbf{g}_k+\sigma_k^2\geq e^{\tau_k},\label{Rs1} \\
&\mathbf{g}_k^H\big(\sum\nolimits_{j\neq k,\;j=0}^K\mathbf{W}_j\big)\mathbf{g}_k+\sigma_k^2\leq e^{\zeta_k},  \label{Rs2}\\
&\mathbf{g}_e^H\big(\sum\nolimits_{k=0}^K\mathbf{W}_k\big)\mathbf{g}_e+\sigma_e^2\leq e^{\tau_e},\label{Rs3} \\
&\mathbf{g}_e^H\big(\sum\nolimits_{j\neq k,\;j=0}^K\mathbf{W}_j\big)\mathbf{g}_e+\sigma_e^2 \geq e^{\zeta_e},  \label{Rs4}\\
 &{\rm log_2}\left(e^{\tau_{k}-\zeta_{k}}\right)-{\rm log_2}\left(e^{\tau_{e}-\zeta_{e}}\right) \geq R_{sk}.\label{Rs5} 
 \end{align}
 \end{subequations}
 \par The constraints \eqref{Rs2} and \eqref{Rs3} remain non-convex due to the exponential terms on the right-hand side. To handle this, we employ a first-order Taylor expansion to approximate the exponential terms and iteratively converge to a stationary point. Specifically, $e^{\zeta_{k}}$ and $e^{\tau_{e}}$ are linearized at the points $\zeta_{k}\left(i\right)$ and $\tau_{e}\left(i\right)$ during the $\left(i+1\right)$-th iteration, where $i$ denotes the iteration index. Accordingly, \eqref{Rs2} and \eqref{Rs3} can be reformulated as
\begin{subequations}
\begin{align}
\begin{aligned}
\mathbf{g}_k^H\sum\nolimits_{j\neq k,\;j=0}^K\!\!\mathbf{W}_j\mathbf{g}_k+\sigma_k^2\!\leq\! e^{\zeta_{k}\left(i\right)}\left[1+\zeta_{k}{\left(i+1\right)}-\zeta_{k}\left(i\right)\right], 
 \end{aligned}\\
 \begin{aligned}
\mathbf{g}_e^H\sum\nolimits_{k=0}^K\!\!\mathbf{W}_k\mathbf{g}_e +\sigma_e^2\!\leq\! e^{\tau_{e}\left(i\right)}\left[1+\tau_{e}{\left(i+1\right)}-\tau_{e}\left(i\right)\right].
\end{aligned}
 \end{align}
 \end{subequations}
 \vspace{-1em}
\par At this point, as the values at the $i$-th iteration can be treated as constants, the originally nonlinear exponential terms can be linearized with respect to the optimization variables in the $\left(i+1\right)$-th iteration. As a result, the SSE constraint is transformed into a tractable convex form. 
\par With the obtained $\mathbf{W}_k$, a feasible rank-one solution can be recovered via the Gaussian randomization method \cite{SDR}, which satisfies all the constraints in the original problem. Our objective is to design the tri-hybrid beamforming architecture such that the overall beamforming matrix can closely approximate the obtained solution $\mathbf{w}_k$. The corresponding optimization problem can be written as
\begin{subequations}\label{P1.2}
\begin{alignat}{2}
\mathcal{P}_{1.2}:\;
&\min_{\mathbf{F}_{\mathrm{dma}}, \mathbf{F}_{\mathrm{RF}}, \mathbf{F}_{\mathrm{dig}}}
&\;&  \Vert\mathbf{F}_{\mathrm{dma}}\mathbf{F}_{\mathrm{RF}}\mathbf{F}_{\mathrm{dig}}-\mathbf{W}\Vert_F^2\label{P1.2:obj}\\
&\hspace{2.7em}\text{s.t.}\;& &  \eqref{Fdma}-\eqref{alphail},\;\eqref{P1:f},
\end{alignat}
\end{subequations}
where we define $\mathbf{W}=[\mathbf{w}_0,\ldots,\mathbf{w}_K]$. Problem $\mathcal{P}_{1.2}$ is a complex optimization problem involving the coupling of three variables. We address it using a triple alternating optimization approach, where each variable is updated sequentially while keeping the others fixed.
\par First, we design the digital beamformer with fixed analog and DMA beamformers. The objective can be transformed into $\min_{\mathbf{F}_{\mathrm{dig}}} \Vert\mathbf{F}_{\mathrm{dma}}\mathbf{F}_{\mathrm{RF}}\mathbf{F}_{\mathrm{dig}}-\mathbf{W}\Vert^2$ without any hardware constraints. This is a typical least-squares (LS) problem, and the theoretically optimal digital beamforming design is
\begin{equation}\label{Fdig}
    \mathbf{F}_{\mathrm{dig}}^{\star}=\big(\mathbf{F}^H_{\mathrm{RF}}\mathbf{F}^H_{\mathrm{dma}}\mathbf{F}_{\mathrm{dma}}\mathbf{F}_{\mathrm{RF}}\big)^{-1}\mathbf{F}^H_{\mathrm{RF}}\mathbf{F}^H_{\mathrm{dma}}\mathbf{W}.
\end{equation}
\par Then, we design the analog beamformer with the resulting digital beamformer in \eqref{Fdig} and fixed DMA beamformer. The sub-problem for the analog beamformer design is
\begin{subequations}\label{P1.3}
\begin{alignat}{2}
\mathcal{P}_{1.3}:\;
&\min_{\mathbf{F}_{\mathrm{RF}}}
&\;&  \Vert\mathbf{F}_{\mathrm{dma}}\mathbf{F}_{\mathrm{RF}}\mathbf{F}_{\mathrm{dig}}^{\star}-\mathbf{W}\Vert^2_F\\
&\text{s.t.}\;& &  \eqref{P1:f}.
\end{alignat}
\end{subequations}
\par Problem $\mathcal{P}_{1.3}$ can be solved through the gradient descent method. Specifically, we define $f(\mathbf{F}_{\mathrm{RF}})=\Vert\mathbf{F}_{\mathrm{dma}}\mathbf{F}_{\mathrm{RF}}\mathbf{F}_{\mathrm{dig}}^{\star}-\mathbf{W}\Vert^2_F$, such that the Euclidean gradient of $f(\mathbf{F}_{\mathrm{RF}})$ with respect to $\mathbf{F}_{\mathrm{RF}}$ is given by
\begin{equation}
\nabla f(\mathbf{F}_{\mathrm{RF}})=2\mathbf{F}_{\mathrm{dma}}^{H}\left(\mathbf{F}_{\mathrm{dma}}\mathbf{F}_{\mathrm{RF}}\mathbf{F}_{\mathrm{dig}}^{\star}-\mathbf{W}\right)\mathbf{F}_{\mathrm{dig}}^{\star H}.
\end{equation}
\par The gradient descent update at the $i$-th iteration is given by
\begin{equation}
{\mathbf{F}}_{\mathrm{RF}}^{(i+1)}
=\mathbf{F}_{\mathrm{RF}}^{(i)}-\mu^{(i)}
\nabla f\left(\mathbf{F}_{\mathrm{RF}}^{(i)}\right),
\end{equation}
where $\mu^{(i)}$ denotes the step size, which is determined by backtracking line search during iteration. Since $\mathbf{F}_{\mathrm{RF}}$ is subject to the constant-modulus constraint in \eqref{P1:f}, ${\mathbf{F}}_{\mathrm{RF}}^{(i+1)}$ is then projected onto the feasible set as
\begin{equation}\label{FRF}
\mathbf{F}_{\mathrm{RF}}^{(i+1)}
=\frac{1}{\sqrt{N_w}}\exp\Big(j\,\angle\big(\mathbf{F}_{\mathrm{RF}}^{(i+1)}\big)\Big),
\end{equation}
where $\angle\big(\mathbf{F}_{\mathrm{RF}}^{(i+1)}\big)$ represents the angle of $\mathbf{F}_{\mathrm{RF}}^{(i+1)}$. The above procedure is repeated until convergence or the maximum number of iterations is reached.

\par After designing the analog beamformer, we fix the analog and digital beamformer and focus on the design of the DMA beamformer. Due to the inherent hardware constraints of the DMA architecture, only the phase shifts are tunable. Thus, the DMA beamformer design can be reformulated as the optimization of the phase variables $\{\theta_{i,l}\}$, which can be expressed as
\begin{subequations}\label{P1.4}
\begin{alignat}{2}
\mathcal{P}_{1.4}:\;
&\min_{\theta_{i,l}}
&\;&  \Vert\mathbf{F}_{\mathrm{dma}}\mathbf{F}_{\mathrm{RF}}\mathbf{F}_{\mathrm{dig}}^{\star}-\mathbf{W}\Vert^2_F\\
&\text{s.t.}\;& & \eqref{Fdma}-\eqref{alphail}.
\end{alignat}
\end{subequations}
\par Problem $\mathcal{P}_{1.4}$ cannot be solved directly due to the hardware constraints. To address this issue, we adopt the coordinate descent (CD) method, where at each step only one phase variable is optimized while keeping the others fixed. For each $\theta_{i,l}$, a one-dimensional (1-D) grid search over $[0,2\pi)$ is performed to find the phase that minimizes the corresponding approximation error. 
Specifically, at each iteration, all phase variables except $\theta_{i,l}$ are fixed, and the original problem can be decomposed into a set of independent 1-D search subproblems. The optimization with respect to $\theta_{i,l}$ can be reformulated as
\begin{equation}\label{Fdmaa}
\theta_{i,l}^{\star}=\arg\min_{\theta_{i,l} \in [0,2\pi)}\sum\nolimits_{k=0}^K\left|m_{i,l}(\theta_{i,l}) z_{i,k} - w_{i,l,k}\right|^2,
\end{equation}
where $w_{i,l,k}$ denotes the element corresponding to the $i$-th radiating element in the $l$-th waveguide of the targeted beamforming matrix $\mathbf{w}_k$, while $z_{i,k}$ represents the $i$-th entry of the equivalent beamforming vector $\mathbf{z}_k=\mathbf{F}_{\mathrm{RF}}\mathbf{f}_{\mathrm{dig}_k}$.
\par The whole triple alternating optimization procedure is iteratively performed, where the DMA, analog, and digital beamformers are alternately updated until the difference between two consecutive fitting errors falls below a predefined threshold $\varepsilon$ or the maximum number of iterations is reached. After optimizing DMA and analog beamformers, the digital beamformer is further refined by solving problem $\mathcal{P}_{1}$ with fixed DMA and analog beamformers, so as to satisfy both SSE and transmit power constraints. The approach for the solution is the same as that used for solving problem $\mathcal{P}_{1.1}$; to avoid redundancy, the details are omitted here. The overall procedure of the proposed triple-alternating algorithm for secure tri-hybrid beamforming design is summarized in \textbf{Algorithm~\ref{al1}}.

\textbf{\textit{Convergence and Complexity Analysis:}} The SCA-based fully-digital optimization converges due to the monotonically non-decreasing and upper-bounded objective. The subsequent procedure  can be regarded as a CD method, where each block update does not increase the approximation error thus guarantees the convergence \cite{b12}. The complexity of using SCA and SDR to solve problem $\mathcal{P}_{1.1}$ is $\mathcal{O}\big(I_{A}(\sqrt{(K+1)N}{\rm log}(\frac{1}{\epsilon_{ipm}})K^2N^3)+I_{G}K^2N^2\big)$, where $I_A$, $I_G$ and $\epsilon_{ipm}$ are the number of SCA iterations, the number of Gaussian randomization, and the accuracy of interior-point method, respectively; the complexity of the triple alternating optimization is $\mathcal{O}\big(I_tNQK\big)$, where $I_t$ and $Q$ denote the number of iterations in the triple alternating algorithm and the number of 1-D phase searches, respectively. %
  \begin{algorithm}[!t] \label{al1}
    \caption{Triple Alternating Algorithm for Secure Tri-hybrid Beamforming Design}
    \LinesNumbered
    \KwIn{$N_{w}, N_r, n_{iter},P, R_{sk},\varepsilon$}
    \KwOut{$\mathbf{F}_{\mathrm{dig}}$, $\mathbf{F}_{\mathrm{RF}}$, $\mathbf{F}_{\mathrm{dma}}$}

    Initialize iteration index $i = 1$, $\mathbf{F}_{\mathrm{dig}}^{(1)}$, $\mathbf{F}_{\mathrm{RF}}^{(1)}$, and $\mathbf{F}_{\mathrm{dma}}^{(1)}$. \\
    Solve problem $\mathcal{P}_{1.1}$ for fully-digital solution $\mathbf{W}$;\\
    Compute the initial approximation error as $J^{(1)}=\Vert\mathbf{F}_{\rm dma}^{(1)}\mathbf{F}_{\rm RF}^{(1)}\mathbf{F}_{\rm dig}^{(1)}-\mathbf{W}\Vert_F^2$.

    \Repeat{$\left|J^{(i)}-J^{(i-1)}\right|\leq \varepsilon$ {\rm or} $i > n_{iter}$}{
    Update $\mathbf{F}_{\rm dig}^{(i+1)}$ by \eqref{Fdig} for fixed $\mathbf{F}_{\rm dma}^{(i)}$ and $\mathbf{F}_{\rm RF}^{(i)}$;\\

    Update $\mathbf{F}_{\rm RF}^{(i+1)}$ by \eqref{FRF} for fixed $\mathbf{F}_{\rm dma}^{(i)}$ and $\mathbf{F}_{\rm dig}^{(i+1)}$;\\

    Update $\mathbf{F}_{\rm dma}^{(i+1)}$ by \eqref{Fdmaa} for fixed $\mathbf{F}_{\rm RF}^{(i+1)}$ and $\mathbf{F}_{\rm dig}^{(i+1)}$;\\

    Set$J^{(i+1)}=\Vert\mathbf{F}_{\rm dma}^{(i+1)}\mathbf{F}_{\rm RF}^{(i+1)}\mathbf{F}_{\rm dig}^{(i+1)}-\mathbf{W}
    \Vert_F^2$;\\
    Set $i \leftarrow i + 1$;}
    Recompute $\mathbf{F}_{\rm dig}$ by resolving problem $\mathcal{P}_{1.1}$;\\
    \Return{$\mathbf{F}_{\mathrm{dig}}$, $\mathbf{F}_{\mathrm{RF}}$, $\mathbf{F}_{\mathrm{dma}}$}.
\end{algorithm}
\setlength{\textfloatsep}{0.1cm}

\section{Numerical Results}
 \par In this section, we provide numerical results to evaluate the performance of the proposed secure tri-hybrid beamforming scheme. The number of UEs is set to $K=2$. The three-dimensional coordinates of the BS, UEs, target, and Eve are set as $[0\;{\rm m},0\;{\rm m},10\;{\rm m}]$, $[-10\;{\rm m},0\;{\rm m},1\;{\rm m}]$, $[10\;{\rm m},0\;{\rm m},1\;{\rm m}]$, $[10\;{\rm m},10\;{\rm m},5\;{\rm m}]$, and $[15\;{\rm m},15\;{\rm m},1\;{\rm m}]$, respectively. The numbers of DMA waveguides and the radiating elements per waveguide are set to $N_w=4$ and $N_r=8$, respectively. The transmit power at the BS is set to $P=3$ Watt, and the carrier frequency is $f_c=28 \;\rm GHz$, corresponding to a wavelength of approximately $\lambda_c=1.07$ cm. The noise variance is set to $\sigma^2=-80$ dBm for all nodes in the system. The spacing between adjacent waveguides is configured to $v=\frac{\lambda}{2}$ and that between adjacent radiating elements is $h=\frac{\lambda}{5}$ \cite{b13}. The attenuation coefficient and wavenumber of DMA are $\alpha=0.6 \rm \;m^{-1}$ and $\beta=827.67\rm \;m^{-1}$. The mmWave channel is modeled as $\mathbf{g}_k=\sqrt{\frac{N}{L_{p}}}\sum_{l=1}^{L_{p}}\alpha_l\mathbf{a}^H\left(\theta_l,\phi_l\right)$, where $\alpha_l$ is the complex channel gain of the \emph{l}-th path, and $L_p$ denotes the total number of paths. In the simulations, we set $L_p=6$.
 
In all simulation results, “\textbf{DMA}” refers to the proposed tri-hybrid beamforming design. The following baseline schemes are considered for comparison:

\begin{itemize}
\item \textbf{FD (SAN):} A fully-digital design with the same number of antennas (SAN) as the proposed scheme.
\item \textbf{HB (SAN):} A hybrid beamforming architecture with the same number of antennas.
\item \textbf{FD (SAA):} A fully-digital design with fewer radiating elements while maintaining the same array aperture (SAA).
\item \textbf{HB (SAA):} A hybrid beamforming counterpart with fewer radiating elements but the same array aperture \cite{b6}.
\item \textbf{DMA (fixed):} A tri-hybrid beamforming scheme with a fixed DMA electromagnetic (EM) response, where the DMA phase shifts are initialized to align with the target direction and remain fixed throughout the optimization, while only the analog and digital beamformers are updated.
\end{itemize}
\vspace{0.4em}
\begin{figure}[t]
    \centering
    \includegraphics[width=2.9in]{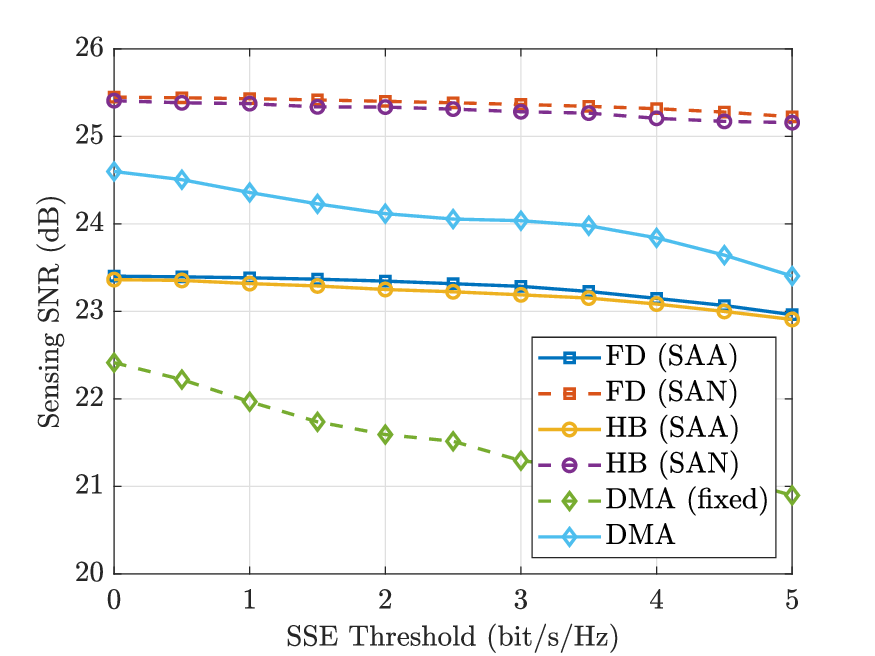}
    \vspace{-0.8em}
    \caption{Sensing SNR versus the SSE threshold.}
    \vspace{-1em}
    \label{fig2}
\end{figure}
 \vspace{-0.5em}
\begin{figure}[t]
    \centering
    \includegraphics[width=2.9in]{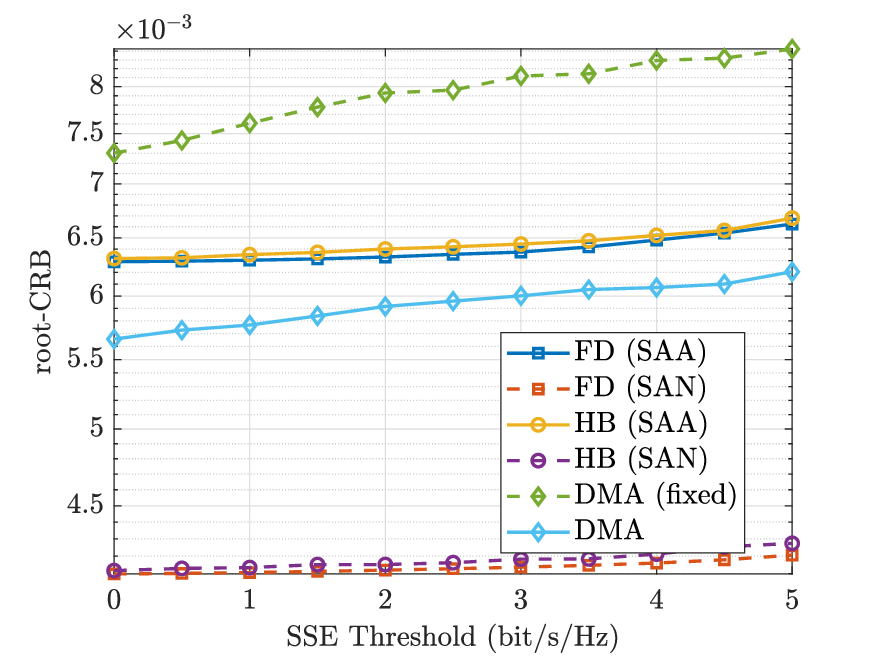}
    \vspace{-0.8em}
    \caption{Root-CRB versus the SSE threshold.}
    \vspace{-0.3em}
    \label{fig3}
\end{figure}
\begin{figure}[!htbp]
    \centering
    \includegraphics[width=2.9in]{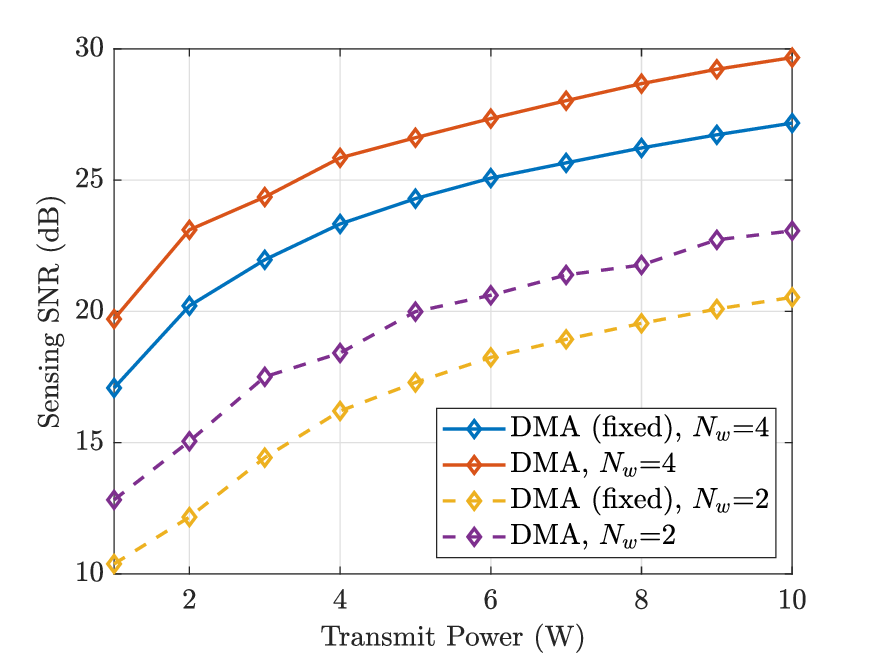}
    \vspace{-0.8em}
    \caption{Sensing SNR versus the transmit power with different $N_w$.}
    \vspace{-0.3em}
    \label{fig4}
\end{figure}
\par Figure 2 shows the sensing SNR as a function of the SSE threshold. As the SSE threshold increases, the BS shall allocate more transmit power and spatial degrees of freedom to the communication beams to satisfy the required secrecy performance. In addition, part of the sensing capability is leveraged to interfere with Eve. Consequently, fewer resources remain available for sensing, leading to a gradual degradation in the sensing performance. Among all considered schemes, the fully-digital beamforming design with the same number of antennas achieves the highest sensing SNR, as its objective directly maximizes sensing performance and each antenna is connected to a dedicated RF chain. This allows maximum design flexibility and precise beam control without structural constraints. Notably, the proposed tri-hybrid beamforming design outperforms both the fully-digital and hybrid beamforming schemes under the same array aperture, and also surpasses the tri-hybrid design with a fixed EM response.

\par To better evaluate the sensing performance, Fig. 3 shows the CRB performance versus the SSE threshold. It is evident that the CRB generally increases as the SSE requirement becomes more strict, because more power needs to be allocated toward the legitimate UEs to ensure secure communication. The proposed tri-hybrid beamforming architecture can achieve a lower CRB than the fully-digital and hybrid architectures with the same aperture size. Note that, compared with the tri-hybrid beamforming design with a fixed EM response, the proposed scheme achieves a greater improvement in CRB performance, indicating that the additional complexity introduced by EM design in the DMA beamformer is justified by the resulting sensing performance gain.
\par Figure 4 illustrates the sensing SNR performance versus the total transmit power with different numbers of waveguides, where we set the threshold of $\SSEk$ to $R_{sk}=1$ bit/s/Hz. As the transmit power increases, the sensing SNR consistently improves since the objective is to maximize the sensing SNR. Moreover, increasing the number of waveguides can allow the transmit beam to be more accurately focused toward the target, thereby further enhancing the sensing SNR.
\section{Conclusion}
\par In this paper, we proposed a secure tri-hybrid beamforming design for DMA-aided ISAC systems, in which the digital, analog, and DMA beamformers were jointly optimized through a triple optimization algorithm. The optimization problem was formulated as maximizing the sensing SNR while constraining on the SSE and the physical limitation of analog and DMA beamformers. To address the complex problem, a fully-digital beamforming solution was first obtained by relaxing the hardware constraints, based on which the tri-hybrid architecture was designed iteratively to approximate the optimized fully-digital solution through a triple alternating algorithm. In addition, the CRB was derived and used to comprehensively evaluate the system performance. Numerical results showed that the proposed secure tri-hybrid beamforming design outperforms the conventional fully-digital and hybrid beamforming design with the same array aperture as well as the tri-hybrid beamforming with fixed EM design.

\end{document}